\documentstyle[bo99,epsfig]{article}

\def\gi{{\it Ginga}}
\def\asca{{\it ASCA}}
\def\etal{{et al.}}

\title{Iron K line emission in AGN: observations}

\author{K.~Nandra $^{1,2}$}
                                                       
\affil{1) NASA/Goddard Space Flight Center, Mail Code 662, Greenbelt, 
MD 20771, USA \\
2) Universities Space Research Association}                                                

\begin{document}

\maketitle

\begin{abstract}
Iron K$\alpha$ lines are key diagnostics of the central regions of
AGN. Their profiles indicate that they are formed deep in the
potential well of the central black hole, where extreme broadening and
red shift occur. The profiles are most easily reproducible in an
accretion disk: the lack of significant emission blue-ward of the rest
energy is difficult produce in other geometries. In one source an
apparent (and perhaps variable) absorption feature in the red wing of
the line may represent rare evidence for inflow onto the black
hole. Sample analysis has defined the mean properties, showing a
strong concentration of the emission in the central regions and
face-on accretion disks, at least in Seyfert 1 galaxies. Surprising
results have been obtained from examination of the line
variability. Strong profile changes may be accounted for by changes in
the illumination pattern of the central, relativistic part of the
disk. In at least the case of MCG-6-30-15, there is evidence for
emission from within $6 R_{\rm g}$, possibly indicating a spinning
black hole.  Developing an understanding of these complex changes has
the potential to reveal the geometry and kinematics of the inner few
gravitational radii around extragalactic black holes.

\end{abstract}

\section{Introduction}
The first iron K$\alpha$ lines were discovered in NGC 4151, and a few
source with large absorbing columns, in which the line was thought to
originate ((Mushotzky, Holt \& Serlemitsos 1978; Mushotzky 1982).  The
first unobscured AGN to show line emission was MCG-6-30-15 (Nandra et
al. 1989; Matsuoka et al. 1990) and \gi\ subsequently found iron
K$\alpha$ emission to be extremely common in Seyfert galaxies (Pounds
et al. 1990; Nandra \& Pounds 1994). Line emission had been predicted
from optically-thick material close to the nucleus (Guilbert \& Rees
1988), including the accretion disk (Fabian et al. 1989).  Detailed
predictions of the line strength from the disk (George \& Fabian 1991;
Matt, Perola \& Piro 1991) were found to be in excellent agreement
with the observations (Nandra \& Pounds 1994), but the \gi\ data were
unable to determine the width or profile of these lines. This is of
clear importance, as the profiles allow the location and geometry of
the material to be constrained.  Specifically, in the case of an
accretion disk, large widths and distinctive profiles are expected due
to the rotation and gravitational effects of the black hole (Fabian et
al. 1989; Stella 1990; Laor 1991; Matt et al. 1992). The launch of
\asca\ offered an opportunity to test these models, with the SIS
detectors having good sensitivity and a factor $\sim 4$ improvement in
energy resolution compared to \gi.

%--------------------------  figure 1
%this section shows how to insert a figure in the text
\begin{figure}
\centerline{\psfig{file=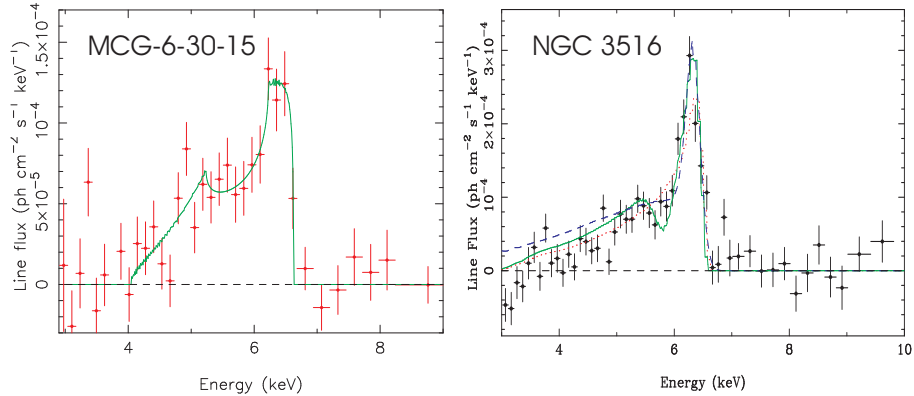, width=12cm}}
\caption[]{Iron K$\alpha$ profiles for MCG-6-30-15 (left panel; Tanaka
et al. 1995) and NGC 3516 (right panel; Nandra et al. 1999). The ASCA
SIS data, derived from interpolating a local continuum, are shown as
the crosses.  The profiles for both sources are very similar, and are
extremely broad. They exhibit a relatively-narrow core peaked at the
rest energy of near-neutral iron ($\sim 6.4$ keV) and a very broad
wing to lower energies. These profiles are characteristic of Doppler
and gravitational effects in an accretion disk orbiting a black hole
and the lines show various models of such emission, which fit the data
extremely well. In NGC 3516, the best model includes an absorption 
line around 5.8 keV. This feature may be due to resonance scattering
in material inflowing into the central regions. 
}
\end{figure}
%---------------------------------

\section{Profiles of individual sources}

The early ASCA data did indeed show evidence that the iron K$\alpha$
lines in AGN were broad with velocity widths of $\sim 50,000$~km
s$^{-1}$, characteristic of material extremely close to the central
black hole (Fabian et al. 1994; Mushotzky et al. 1995).  Uncertainties
in calibration, limited sensitivity above $\sim 7$~keV and the short
exposures of these early observations made it difficult to determine
the profiles, however. Long observations have provided the best
constraints:

\begin{table}
\begin{tabular}{lll}
\hline
Parameter        & Symbol & Typical range \\
\hline
\hline
Rest Energy      & E & 6.4-6.9 keV  \\
Inclination      & i & 0-90 degree \\
Inner radius     & $R_{\rm i}$ & 1-6 $R_{\rm g}$ \\
Outer radius     & $R_{\rm o}$ & 20-1000 $R_{\rm g}$ \\
Emissivity Index & $q$         & 0-3 \\
Equivalent Width & $EW$        & 100-500 eV \\
\hline
\end{tabular}
\caption{Disk line parameters}
\end{table}

{\bf MCG-6-30-15:} The first high signal-to-noise ratio profile was
obtained in a $\sim 150$~ks observation of the Seyfert 1 galaxy
MCG-6-30-15 (Tanaka et al. 1995). These data provided strong
confirmation of the hypothesis that the iron K$\alpha$ line arises
from the inner accretion disk (Fig.~1). The profile is extremely
broad, with FWZI$\sim 0.3$c, and is skewed to the red. This is
characteristic of accretion disk models in which the disk is observed
close to face-on, where the dominant broadening process is the strong
gravitational field of the black hole, rather than Doppler
motions. Indeed, the particular characteristics of the line in
MCG-6-30-15 are extremely difficult to explain without the invocation
of a black hole and accretion disk (see, e.g. Fabian et al. 1995). The
line profile of MCG-6-30-15 has therefore rightly received much
attention and scrutiny, as it presents arguably the most direct
evidence we have for the existence of black holes in active
galaxies. The broad profile in MCG-6-30-15 has been confirmed by
BeppoSax (Guainazzi et al. 1999).

{\bf NGC 4151:} Yaqoob \etal\ (1995) presented a profile very similar
to MCG-6-30-15 for NGC 4151, based on another long ASCA exposure. In
this case the origin of the line is less clear-cut, as the complexity
of the continuum in NGC 4151 makes the line difficult to model
(Zdziarski et al. 1996).  Nonetheless, the similarity of the two
profiles is highly suggestive of a common origin.

{\bf NGC 3516:} A third example (Nandra et al. 1999; Fig. 1), which
again shows a profile remarkably similar to MCG-6-30-15. Once more, an
origin in a face-on accretion disk orbiting a black hole is indicated
and in this case there is also evidence for an absorption feature in
the red wing of the line. This feature may be due to resonance
scattering by iron, redshifted from the rest energy.  If the redshift
is due to kinematic effects, this feature presents rare evidence for
material inflowing into the black hole and could be an important
tracer of accretion. The interpretation is not unique, however, as it
is possible that redshift is gravitational, in which case it indicates
that there may be an ionized ``skin'' above or around the accretion
disk. Ruzkowski \& Fabian (these proceedings) show that this also fits
the data.

\section{Accretion disk models}

Fig.~1 shows various models of line emission from an accretion disk,
which are used to fit the ASCA data. The disk line models such as
those of Fabian et al. (1989) and Laor (1991) are characterized by a
number of parameters, which can in principle be constrained by the
data (Table 1). It has already been mentioned that the very red
profiles of MCG-6-30-15 and NGC 3516 favor low inclinations for the
disk.  Because the emission tends to be centrally-concentrated, the
inner radius of the disk is usually better constrained than its outer
radius. The former is of particular interest because it can help
constrain the black hole spin.  The innermost stable orbit around the
black hole metric: for a Schwarzschild (non-rotating) hole this occurs
at 6 $R_{\rm g}$; for a rapidly rotating hole, however, the stable
orbits exist close to the gravitational radius.  Another crucial
parameter is the line emissivity law, which parameterizes the X-ray
illumination of the disk.  In the models of Fabian et al. (1989) and
Laor (1991), a power-law emissivity assumed, which is a useful
parameterization, if somewhat unphysical.  The true emissivity depends
on the geometry of the X-ray source and accretion disk and their
relationship, modified by relativistic effects and ionization. As the
geometries are very poorly known, what is really required is to
formulate specific physical models and compare them with the
data. Alternatively, one can attempt to ``invert'' the problem and
derive the emissivity from the line profile (Dabrowski et al. 1997;
Cadez \& Calvani, these proceedings).

%--------------------------  figure 2
%this section shows how to insert a figure in the text
\begin{figure}[t]
\centerline{\psfig{file=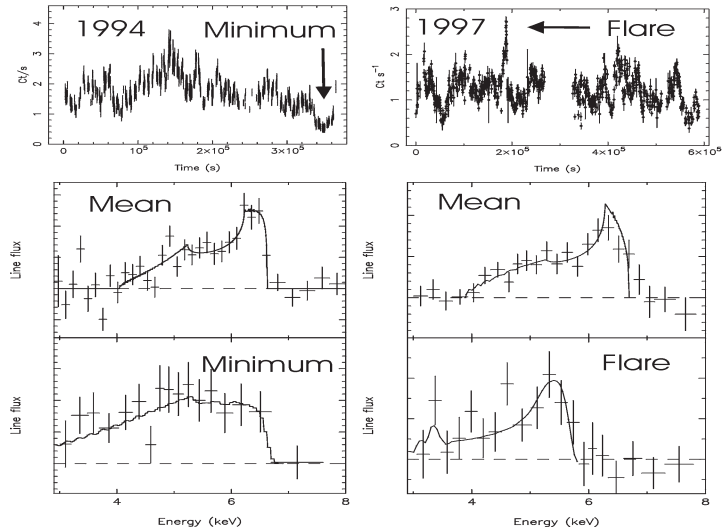, width=12cm, height=8cm}}
\caption[]{
The top panels shows the light curves of MCG-6-30-15 from the long
ASCA observations in 1994 and 1997 (Iwasawa et al. 1996, 1999). The
integrated line profiles of both observations are very similar (middle
panels). In both observations, however, the line profile was found
to vary. In the earlier data, a very broad and redshifted profile was
observed during a deep minimum in the flux. In 1997, a similarly
extreme profile was observed, but this time during a flare. The profile
variations may be attributed to changes in the illumination pattern of the
disk due to localized flares. 
}
\end{figure}
%---------------------------------

\section{AGN Samples}

The profiles of individual sources provide strong constraints, but
studying samples has also been extremely informative. The widespread
applicability of the disk line models has been demonstrated by the
fact that they fit the data better than symmetric profiles, such as a
gaussian (Nandra et al. 1997a hereafter N97; Reynolds 1997).  The
parameters can also be constrained.  

N97 presented ASCA iron K$\alpha$ data for 18 Seyfert 1 galaxies and
found good constraints on the inclination of a number of these. Low
inclinations are very strongly preferred, with a mean of $\sim
30\deg$. As these are Seyfert 1 galaxies, this may not be considered
surprising, as in standard unification schemes, highly inclined
sources would be expected to be obscured, and seen as Seyfert 2
galaxies (Lawrence \& Elvis 1982; Antonucci \& Miller 1995). It is
puzzling then, that Turner et al. (1998) also found low inclinations
to be preferred for a sample of Seyfert 2s and NELGs. A possible
solution to this is suggested by Weaver \& Reynolds (1998), who fitted
the spectra with an additional, narrow line at 6.4 keV, presumed to
arise from the obscuring torus (Ghisselini, Haardt \& Matt 1994;
Krolik, Madau \& Zycki 1994). If such a line is allowed, then so is a
higher inclination of $\sim 50\deg$.

%--------------------------  figure 3
%this section shows how to insert a figure in the text
\begin{figure}[t]
\centerline{\psfig{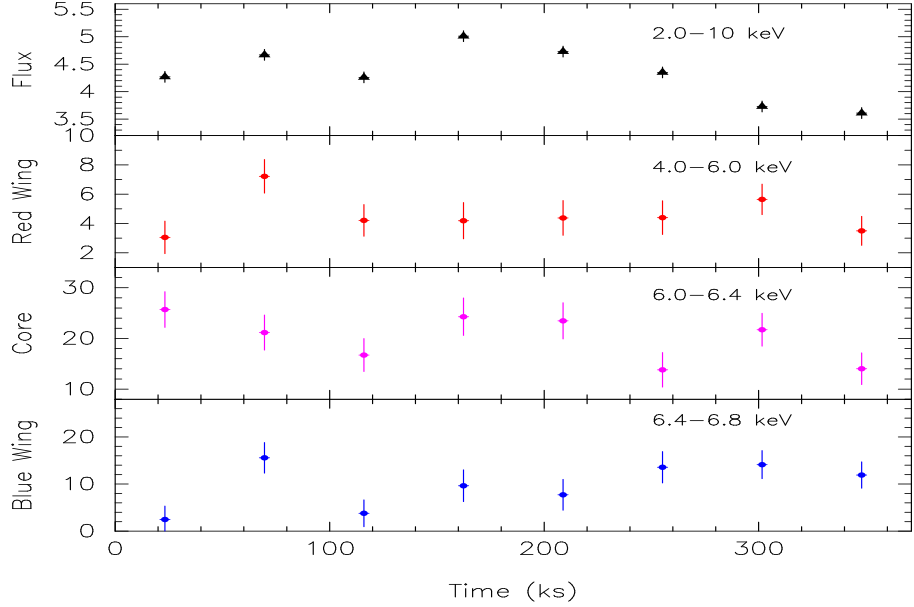}}
\caption[]{Light curves of NGC 3516. In descending order they
are the 2-10 keV continuum, and the excess flux above the continuum in
three line bands.  Neither the core nor the blue wing flux is
consistent with a constant and though the red wing is
formally consistent with no variability, it appears strongly
correlated (at 95\% confidence) with the blue wing.

}
\end{figure}
%---------------------------------

The mean emissivity was found by N97 to be proportional to $R^{-2.5}$
for Seyfert 1 galaxies. This rather steep function implies that the
illumination - and therefore the line emission - is strongly
concentrated in the inner regions of the AGN, with $\sim 50$~\% of the
line coming from within $20 R_{\rm g}$ and 80 \% from within
$100 R_{\rm g}$.  Such an emissivity is roughly consistent with that
of a centrally-illuminated disk.  No universal emissivity law was
found, however, suggesting that there is no single geometry.  

The emission line usually peaks at 6.4 keV, and for low-inclination
disks this implies a low state of ionization, typically $<$Fe {\sc
xx}. If the peak is due to a contribution from a narrow line from
another source, however, no constraint can be placed on the line rest
energy (N97), and therefore ionization.  Standard accretion disks are
expected to be very dense, and can therefore remain cool despite the
intense illumination. Nonetheless, significant ionization might be
expected especially in the central regions, where the X-rays are most
strongly concentrated (e.g. Matt, Fabian \& Ross 1993). A range of
ionization sates throughout the disk is certainly consistent with the
current data. The inner disk could be highly ionized, with the
gravitational and Doppler shifts dominating and making it difficult to
tie down the rest energy.  The observed profiles are well-fit with a
disk line alone, and in Seyfert 1s there is no strong requirement from
line emission from other regions (N97).  In Seyfert 2s, however, there
is evidence for an additional component (Weaver \& Reynolds 1998) and
it is quite plausible that the optical BLR or obscuring torus could
contribute to the line emission in all AGN. The difficulty is
in distinguishing such emission from that of the low-velocity outer disk,
which requires high spectral resolution.

\subsection{Differences and similarities}

All of the high quality profiles obtained so far show profiles which
are remarkably similar, and similar to the composite profile of
Seyfert 1s. There are a few cases where it has been claimed that the
line is narrow (e.g. NGC 4051 Mihara et al. 1994; NGC 7469 Guainazzi
et al. 1994; Mrk 766 Leighly et al. 1996) but it is very difficult to
exclude a broad component with the ASCA data, and none of these case
is clear cut. Nonetheless there do appear to be differences in the
profiles comparing different sources.  The lack of a universal
emissivity law is one suggestion of this (N97). Good quality data for
more sources is required to confirm this, and investigate the origin.
They may, for example, represent differences in geometry. Alternatively, 
they could be as simple as variations in the relative contributions of 
narrow and disk-line components.

\section{Variability studies}

Variability studies often contribute fundamentally to our understanding of
AGN. The line emission is no exception, and here some of the observations
are reviewed.

\begin{itemize}

\item{\bf MCG-6-30-15:} During the observation reported by Tanaka et
al. (1995) the profile of the line was found to be variable
(Fig. 3). Iwasawa et al. (1996) reported an unusually broad and
redshifted profile during a ``deep minimum'' in continuum
flux. Another long observation in 1997 showed a similar mean profile
but this time exhibited the extreme broadening during a flare (Iwasawa
et al. 1999). The profile variability was interpreted as being due to
changes in the illumination pattern of the disk. If, instead of a
single coherent source, localized flares produce the X-rays, then at
certain times a few or even a single flare could dominate the
emission. If that flare occurred in the very innermost regions, the
line profile would temporarily appear more redshifted than the average.
In MCG-6-30-15 the line is so broad during the deep minimum
that it implies that the line emission arises within $\sim 6$~R$_{\rm
g}$ thus implying a Kerr black hole (Iwasawa et al. 1996; see also
Reynolds \& Begelman 1997; Dabrowski et al. 1997; Young et al. 1999).

\item{\bf NGC 3516:} Nandra et al. (1999) have noted profile
variability in NGC 3516 also (Fig.~3). The absorption feature at 5.8
keV (Fig. 1) was found to be prominent only in the middle part of the
observation, where the flux was high. Furthermore, while the core of
the line appeared to follow the variations in the continuum, the red
and blue wings wings did not. They were well correlated with each
other, however, and showed show strong variability (factor $\sim 2$).
This was in excess of the variations in the driving continuum and
therefore very difficult to explain in standard models. An
interpretation similar to MCG-6-30-15 remains valid, where a localized
flare beamed towards the inner disk causes the
variation. Alternatively, or additionally, flares could cause an
increase in the ionization of the inner disk. This could cause on
over-response because the effective fluorescence yield increases
sharply in the helium-like and hydrogen-like ionization states.
Nandra et al. (1997b) also noted significant variability of the line
flux in NGC 3516 over a 1 year baseline, with no obvious change in
profile.

\item{\bf Other sources:} Yaqoob et al. (1996) found evidence for
rapid variability of the broad wing in \underline{\it NGC 7314} but no
obvious changes were seen in the core. This is consistent with the
simple disk model, where the red wing is expected to come from close
in. In \underline{\it NGC 4051} Wang et al. claimed a significant flux
change comparing two ASCA observations, though some portion of the
line flux is constant and comes from a more distant region (Uttley et
al. 1999). Chiang et al. (1999) found no evidence for changes in the
line flux of \underline{\it NGC 5548} despite large variability of the
continuum.

\end{itemize}

\section{Problems and open issues}

Some of the iron K$\alpha$ line observations - particularly in the
realm of variability - have been surprising in the context of the
standard disk-line model. Depending on ones viewpoint, this may be
interpreted as a further demonstration of their diagnostic power (see
above) or as being problematic for the disk line model (e.g.,
Sulentic, Marziani \& Calvani 1998a). Alternatives to the relativistic
disk have been suggested and we review those briefly here. Many of
these issues have been discussed by Fabian et al. (1995). They
include:

{\it Alternative geometries:} in the original paper predicting the
iron K$\alpha$ lines, Guilbert \& Rees (1988) suggested that they may
come from optically thick clouds, sheets or filaments, rather than the
disk. Such material can produce the observed line strengths, as long
as the covering fraction is high, but less than unity to avoid
obscuring the continuum (e.g. Bond \& Matsuoka 1993; Nandra \& George
1994). Detailed calculations of the profiles in a pseudo-spherical
geometry have not been carried out and it is therefore difficult to
make a definite statement as to whether the observations can be
reproduced. It does, however, seem unlikely. In MCG-6-30-15 and NGC
3516 (Fig.~1), there is very little evidence for emission blue-ward of
the rest energy, and it is hard to envisage a non-disk geometry in
which the blue emission is suppressed. The details depend on the
geometry, kinematics and self-covering of the cloud system but, for
example rotating blobs are almost certainly excluded, given that the
high velocities would cause Doppler boosting of the blue
wing. Inflowing or outflowing cloud distributions are another
possibility, but once again it seems hard to suppress the emission
from material moving towards us without a special geometry.

{\it Complex continuum:} As mentioned above, Zdziarski et al. (1996)
have suggested that an apparently-broad line in NGC 4151 can be
accounted for by a complex continuum - in particular complex
absorption. Many AGN have material in their lines-of-sight, such as
``warm absorbers'' (e.g. Reynolds 1997; George et al. 1998) but
typically these are not thought to affect the spectrum above about 3
keV. As in all other wavebands, a great challenge in determining the
line strengths and profiles is determining the underlying
continuum. Although different continuum-deconvolution methods tend to
produce similar results, unknown complexities add additional
uncertainty to the profile determination.

{\it Comptonization:} Kinematics and gravitation are not the only
mechanisms by which lines can be broadened. A particular alternative
suggestion has been that Compton scattering can broaden the line
(e.g., Misra \& Khembavi 1998). Fabian et
al. (1995) argued against such a model, as the Comptonizing medium
must be finely-tuned to generate the observed profile. In particular a
very compact, high optical depth ($\tau \sim 5$), cool ($\sim
0.2$~keV) medium is required. Such a medium would down-scatter the
X-ray continuum also, producing a spectral break at $\sim 20$~keV
which is not observed.

{\it Line blending:} Combinations of numerous lines can appear broad
when observed at low resolution. Differently-ionized species of
iron-K, for example, result in emission from 6.4-6.9 keV.  This cannot
explain the observed profiles, as the broad emission occurs below 6.4
keV. No abundant element produces line emission in the 4-6 keV region
where the red wing is observed, but Skibo (1997) has suggested that
spallation of iron nuclei by $> 10$~MeV photons could result in
significant amounts of V, Ti, Cr and Mn. There are numerous
difficulties with such a model. For example, the observed profile
changes (see above) argue forcefully against such a suggestion, as all
the fluorescence lines should vary together.
 
\subsection{Open issues}

None of these alternatives is as compelling as the standard disk, but
some open questions remain. The broad lines have been confirmed by
BeppoSax (Guainazzi et al. 1999), but we await further confirmation
and definition of the profiles.  In particular, it is important to
quantify and deconvolve the disk line contribution from narrower
components from the BLR and obscuring torus. ASCA observations have
shown some differences comparing objects, but the high signal-to-noise
profiles all show common features, and in particular the derived
inclinations are often very similar. This, and possible disagreements
with other inclination indicators (Sulentic et al. 1998b) are not yet
significant problems, but it will be important in the future to relate
the properties derived from iron K$\alpha$ with other AGN observables.
With large collecting area, it may be possible to discover the weak
and very broad lines expected from edge-on accretion disks, which have
so far eluded us. Deconvolution from a (potentially-complex) continuum
is the key here.

The K$\alpha$ observations so far have been of great importance, but
have shown unexpected complications. The interpretation of future
observations is therefore likely to be challenging. Two particular
issues are the emissivity and ionization of the disk as a function of
radius. These are both arbitrary from an observational standpoint, and
difficult to predict theoretically. Detailed interpretation of, e.g.,
variability data - including reverberation mapping (e.g. Reynolds et
al. 1999) - will require an understanding of these effects.  Iron
K$\alpha$ observers can therefore look forward to developing the kind
of complex physical models and advanced data analysis techniques that
many other astronomers have been enjoying for many years. We hope and
expect the rewards to be substantial.

\begin{acknowledgements}
I am grateful to Andy Fabian, Ian George, Kazushi Iwasawa, Richard
Mushotzky, Chris Reynolds, Jane Turner and Tahir Yaqoob for much
discussions and data. Financial support is provided by NASA grant NAG
5-7067, through USRA.
\end{acknowledgements}

\end{document}